\documentclass[pra,twocolumn,aps,superscriptaddress,showpacs]{revtex4-1}

\usepackage{hyperref}
\usepackage{graphicx}
\usepackage{amsmath}
\usepackage{amsfonts}
\usepackage{amssymb}
\usepackage{lineno}
\usepackage{epsfig}
\usepackage{subfigure}
\usepackage[usenames,dvipsnames]{color}
\usepackage{setspace}
\usepackage{bm}
\usepackage{times}
\hypersetup{
      colorlinks=true,
      citecolor=blue,
      linkcolor=blue,
      urlcolor=blue}

\begin{document}
\title{Characteristic temperature for the immiscible-miscible transition of
       binary condensates in optical lattices
       }

\author{K. Suthar}
\affiliation{Physical Research Laboratory,
             Navrangpura, Ahmedabad-380009, Gujarat,
             India}
\affiliation{Indian Institute of Technology,
             Gandhinagar, Ahmedabad-382424, Gujarat, India}

\author{D. Angom}
\affiliation{Physical Research Laboratory,
             Navrangpura, Ahmedabad-380009, Gujarat,
             India}

\date{\today}


\begin{abstract}
 
 We study a two-species Bose-Einstein condensates confined in 
quasi-two-dimensional (quasi-2D) optical lattices at finite temperatures, 
employing the Hartree-Fock-Bogoliubov theory with the Popov approximation. We 
examine the role of thermal fluctuations on the ground-state density 
distributions, and the quasiparticle mode evolution. At zero temperature, the 
geometry of the ground-state in the immiscible domain is side-by-side. Our 
results show that the thermal fluctuations enhance the miscibility of the 
condensates, and at a characteristic temperature the system becomes miscible 
with rotationally symmetric overlapping density profiles. This 
immiscible-miscible transition is accompanied by a discontinuity in the 
excitation spectrum, and the low-lying quasiparticle modes such as slosh mode 
becomes degenerate at the characteristic temperature.  

\end{abstract}



\maketitle


\section{Introduction}
 Ultracold atoms in an optical lattice offer fascinating prospects to 
investigate many-body quantum physics of strongly correlated systems in a 
highly controllable environment~\cite{jaksh_98,orzel_01,greiner_02,bloch_12}. 
These systems are recognized as ideal tools to explore new quantum 
phases~\cite{demler_02,kuklov_03,kuklov_04}, 
complex phase transitions~\cite{pal_10,sungsoo_11,lin_15,jurgensen_15}, quantum
magnetism~\cite{trotzky_08,simon_11}, quantum information~\cite{bloch1_08} and 
to simulate transport and magnetic properties of condensed-matter 
systems~\cite{lewenstein_07,bloch_08}. Moreover, the effect of phase 
separation~\cite{mishra_07,zhan_14}, quantum emulsions and coherence 
properties~\cite{greiner_01,roscilde_07,buonsante_08}, and multicritical 
behaviour~\cite{ceccarelli_15,ceccarelli_16} of the mixtures have been explored
in the past decade. 

Among the various observations made in the two-species Bose-Einstein 
condensates (TBECs) of ultracold atomic gases, the most remarkable is the 
phenomenon of phase separation, and it has been a long-standing topic of 
interest in chemistry and physics. For repulsive on-site interactions, the 
transition to the phase-separated domain or immiscibility is characterized by 
the parameter $\Delta = U_{11} U_{22}/U^{2}_{12} - 1$, where $U_{11}$ and 
$U_{22}$ are the intraspecies on-site interactions and $U_{12}$ is the 
interspecies on-site interaction. When $\Delta < 0$, an immiscible phase occurs
in which, the atoms of species $1$ and $2$ have relatively strong repulsion, 
whereas $\Delta\geqslant 0$ implies a miscible 
phase~\cite{ho_96,timmermans_98,esry_99}. It is important to note that the 
mention criterion is valid at zero temperature for homogeneous systems. The 
presence of an external trapping potential, however, modifies this condition as
the trap introduces an additional energy cost for the species to spatially 
separate~\cite{wen_12}. In experiments, the unique feature of phase separation 
has been successfully observed in TBECs with harmonic trapping 
potential~\cite{papp_08,tojo_10,mccarron_11}. Previously, in the context of 
superfluid Helium at zero temperature, the phase separation of the bosonic 
mixtures of isotopes of different masses has also been predicted in 
Refs.~\cite{chester_55,miller_78}. The recent experimental realizations of 
TBECs in optical lattices, either of two different atomic 
species~\cite{catani_08} or two different hyperfine states of same atomic 
species~\cite{gadway_10,soltan_11} provide the motivation to study these 
systems in detail. In recent works, we have examined the miscible-immiscible 
transition, and the quasiparticle spectra of the TBECs at zero temperature in 
quasi-1D~\cite{suthar_15} and quasi-2D~\cite{suthar_16} geometries. The finding
in the latter work~\cite{suthar_16}, where we had examined the nature of the 
density profiles in immiscible regime at zero temperature, is of relevance to 
the present work. In addition, we had shown how the optical lattice potential 
influences the density profiles in the immiscible domain. The other related 
study is the ground-state phase diagram, and the effect of filling factor of 
the TBECs on the phenomenon of phase separation, which were investigated using 
quantum Monte Carlo simulations~\cite{lingua_15,galteland_15}. In addition, 
phase-separation of TBECs at various length scales has been examined using the 
multi-orbital mean-field theory~\cite{alon_06,alon_07b}. Among the full quantum
methods the multi-configurational time-dependent Hartree for bosons (MCTDHB) 
provide a good description of the formation of the interference fringes in the 
densities during the mixing of condensates~\cite{alon_07,alon_08}. This method 
allows the dynamical creation of quantum superposition of states in ultracold 
Bose gases~\cite{cederbaum_07}. In other theoretical studies, the finite 
temperature properties of TBECs have 
been explored~\cite{ohberg_99,shi_00,kwangsik_07}. In continuum or TBECs with 
harmonic confining potential alone, we have explored the suppression of phase 
separation due to the presence of the thermal fluctuations~\cite{arko_15}. 
However, a theoretical understanding of the finite temperature effects on the 
topology and the collective excitations of TBECs in optical lattices is yet to 
be explored. The Bose-Einstein condensation and hence, the coherence in a 
system of bosons depends on the interplay between various parameters, such as 
temperature, interaction strength, confinement, and 
dimensionality~\cite{proukakis_06}. In particular, in the low-dimensional Bose 
gases, the coherence can only be maintained across the entire spatial extent at
a temperature much below the critical temperature. The coherence property, in 
experiments, have been studied in recent 
works~\cite{dettmer_01,hellweg_03,richard_03,esteve_06,plisson_11}.

With an attention towards this unexplored physics, we study the finite 
temperature effects of quasi-2D trapped TBECs in optical lattices. In the 
present work, we address the topological phase transition in the TBECs of two 
different isotopes of Rb with temperature as a control parameter in the domain 
$T<T_c$, where $T_c$ is the critical temperature of either of the species of 
the mixture. Here, it must be mentioned that in our previous 
works~\cite{suthar_15,suthar_16}, we had investigated the ground-state density,
and the quasiparticles with variation in on-site interaction energy at zero 
temperature. In addition, we have examined the effect of quantum fluctuations 
on the ground state geometry and collective excitations of the quasi-1D TBECs. 
In the present work, we shall examine the evolution of the quasiparticle modes 
of TBECs in quasi-2D optical lattices with variation in temperature. For this 
work, we use Hartree-Fock-Bogoliubov (HFB) formalism with the Popov 
approximation, and starting from phase-separated domain at zero temperature we 
increase temperature. We observe that there is an immiscible to miscible 
transition of the TBEC at a characteristic temperature. This transition is 
accompanied by a discontinuity in the quasiparticle excitation spectrum, and in
addition, some of the modes like the slosh mode become degenerate. We, then, 
compute the equal-time first-order spatial correlation function which is a 
measure of the coherence and phase fluctuations present in the system. It 
describes the off-diagonal long range order which is a defining feature of 
BEC~\cite{penrose_56}. This is an important theoretical tool to study the many 
body effects in atomic physics experiments~\cite{burt_97,tolra_04}. 

This paper is organized as follows. In Sec.~\ref{theory_2s2d} we describe the 
HFB-Popov formalism, and the numerical techniques used in the present work. 
The evolution of the quasiparticle modes and the density distributions with 
the temperature are shown in Sec.~\ref{results}. Finally, our main results 
are summarized in Sec.~\ref{conc}.


\section{Theory and methods}
\label{theory_2s2d}
\subsection{HFB-Popov approximation for quasi-2D TBEC}
 We consider a TBEC confined in an optical lattice with pancake-shaped 
configuration of background harmonic trapping potential. Thus, the trapping
frequencies satisfy the condition $\omega_{\perp} \ll \omega_z$ with 
$\omega_x = \omega_y = \omega_{\perp}$. In this system, the excitation energies
along the axial direction are high, and the degree of freedom in this direction
is frozen. The excitations, both the quantum and thermal fluctuations, are 
considered only along the radial direction. In the tight-binding approximation 
(TBA)~\cite{chiofalo_00,smerzi_03}, the Bose-Hubbard (BH) 
Hamiltonian~\cite{fisher_89,lundh_12,hofer_12} describing this system is
\begin{eqnarray}
\hat{H} = && \sum_{k=1}^2 \bigg[- J_k \sum_{\langle \xi\xi'\rangle} 
             \hat{a}^{\dagger}_{k\xi}\hat{a}_{k\xi'} 
           + \sum_\xi(\epsilon^{(k)}_{\xi} - \mu_k)
              \hat{a}^{\dagger}_{k\xi}\hat{a}_{k\xi}\bigg] \nonumber\\
          &+& \frac{1}{2}\!\!\sum_{k=1, \xi}^{2}\!\! 
              U_{kk}\hat{a}^{\dagger}_{k\xi}
              \hat{a}^{\dagger}_{k\xi}\hat{a}_{k\xi}\hat{a}_{k\xi} 
          + U_{12}\!\!\sum_\xi \hat{a}^{\dagger}_{1\xi}\hat{a}_{1\xi}
              \hat{a}^{\dagger}_{2\xi}\hat{a}_{2\xi},
\label{bh2d}              
\end{eqnarray}
where $k = 1,2$ is the species index, $\mu_k$ is the chemical potential of the 
$k$th species, and $\hat{a}_{k\xi}$ ($\hat{a}^\dagger_{k\xi}$) is the 
annihilation (creation) operators of the two different species at $\xi$th 
lattice site. The index is such that $\xi \equiv (i,j)$ with $i$ and $j$ as the
lattice site index along $x$ and $y$ directions, respectively. The summation 
index $\langle \xi\xi'\rangle$ represents the sum over nearest-neighbour to the
$\xi$th site. The TBA is valid when the depth of the lattice potential is much 
larger than the chemical potential $V_0 \gg \mu_k$, the BH Hamiltonian then 
describes the system when the bosonic atoms occupy the lowest energy band. A 
detailed derivation of the BH Hamiltonian is given in our previous 
works~\cite{suthar_15,suthar_16}. In the BH Hamiltonian, $J_k$ are the 
tunneling matrix elements, $\epsilon^{(k)}_{\xi}$ is the offset energy arising 
due to background harmonic potential, and $U_{kk}$ ($U_{12}$) are the 
intraspecies (interspecies) interaction strengths. In the present work all the 
interaction strengths are considered to be repulsive, that is, 
$U_{kk},U_{12}>0$.

In the weakly interacting regime, under the Bogoliubov 
approximation~\cite{griffin_96,amrey_04}, the annihilation operators at each 
lattice site can be written as 
$\hat{a}_{1\xi} = (c_{\xi} + \hat{\varphi}_{1\xi})e^{-i \mu_1 t/\hbar}$, 
$\hat{a}_{2\xi} = (d_{\xi} + \hat{\varphi}_{2\xi})e^{-i \mu_2 t/\hbar}$,
where $c_{\xi}$ and $d_{\xi}$ are the complex amplitudes describing the 
condensate phase of each of the species. The operators $\hat{\varphi}_{1\xi}$ 
and $\hat{\varphi}_{2\xi}$ representing the quantum or thermal fluctuation part
of the field operators. Further more, we consider the system in the superfluid 
domain where the mean-field description is applicable, and accordingly, the 
parameters satisfy the condition 
$U/J\ll 16.7$~\cite{elstner_99,wessel_04,capogrosso_08}. In this domain, the 
equation of motion of the condensate in optical lattice with TBA is reduced to 
discrete nonlinear Schr\"odinger equation (DNLSE). However, in the 
Mott-insulator phase, $U/J\geqslant 16.7$, the mean-field description breaks 
down, and a full quantum description is 
required~\cite{gerbier_07,toth_11,klaiman_17}. From the equation of motion of 
the field operators with the Bogoliubov approximation, the equilibrium 
properties of a TBEC is governed by the coupled generalized DNLSEs
\begin{subequations}
 \begin{eqnarray}
  \mu_1 c_\xi = &-& J_1 \sum_{\xi'} c_{\xi'} + \left [\epsilon^{(1)}_\xi 
               + U_{11} (n^{c}_{1\xi} + 2 \tilde{n}_{1\xi}) 
                + U_{12} n_{2\xi} \right ]  c_\xi,
                      \nonumber \\~\\
  \mu_2 d_\xi = &-& J_2 \sum_{\xi'} d_{\xi'} + \left [\epsilon^{(2)}_\xi 
               + U_{22} (n^{c}_{2\xi} + 2 \tilde{n}_{2\xi}) 
                 + U_{12} n_{1\xi} \right ] d_\xi, 
                      \nonumber \\
 \end{eqnarray}
 \label{dnls2d}
\end{subequations}
where $n^{c}_{1\xi} = |c_\xi|^2$ and  $n^{c}_{2\xi} = |d_\xi|^2$, 
$\tilde{n}_{k\xi} 
= \langle {\hat{\varphi}}^{\dagger}_{k\xi}\hat{\varphi}_{k\xi} \rangle$, and 
$n_{k\xi} = n^{c}_{k\xi} + \tilde{n}_{k\xi}$ are the condensate, noncondensate,
and total density of the species, respectively. The fluctuation operators are 
defined in terms of the quasiparticles through the Bogoliubov transformation
\begin{equation}
  \hat\varphi_{k\xi} = \sum_l\left[u^l_{k\xi}\hat{\alpha}_l e^{-i \omega_l t} 
             - v^{*l}_{k\xi}\hat{\alpha}^{\dagger}_l e^{i \omega_l t}\right],
 \label{bog_trans_2d}                        
\end{equation}
where $\hat{\alpha}_l (\hat{\alpha}^{\dagger}_l)$ are the quasiparticle 
annihilation (creation) operators, which satisfy the Bose commutation 
relations, $l$ is the quasiparticle mode index, $u^l_{k\xi}$ and  $v^l_{k\xi}$ 
are the quasiparticle amplitudes for the $k$th species, and 
$\omega_l = E_l/\hbar$ is the frequency of the $l$th quasiparticle mode with 
$E_l$ as the mode excitation energy.

Using the Bogoliubov transformation, we obtain the following HFB-Popov 
equations~\cite{suthar_16}:
\begin{subequations}
 \begin{eqnarray}
  E_l u^l_{1,\xi} = &-& J_1(u^l_{1,\xi-1} + u^l_{1,\xi+1}) 
                        + \mathcal{U}_1 u^l_{1,\xi} 
                        - U_{11} c^2_\xi v^l_{1,\xi} \nonumber\\ 
                    &+& U_{12} c_\xi(d^{*}_\xi u^l_{2,\xi} 
                        - d_\xi v^l_{2,\xi}),\\
  E_l v^l_{1,\xi} = &~& J_1(v^l_{1,\xi-1} + v^l_{1,\xi+1}) 
                        + \underline{\mathcal{U}}_1
                        v^l_{1,\xi} + U_{11} c^{*2}_\xi u^l_{1,\xi} \nonumber\\
                    &-& U_{12} c^{*}_\xi(d_\xi v^l_{2,\xi} 
                        - d^{*}_\xi u^l_{2,\xi}),\\
  E_l u^l_{2,\xi} = &-& J_2(u^l_{2,\xi-1} + u^l_{2,\xi+1}) 
                        + \mathcal{U}_2 u^l_{2,\xi} 
                        - U_{22} d^2_\xi v^l_{2,\xi} \nonumber\\ 
                    &+& U_{12} d_\xi(c^{*}_\xi u^l_{1,\xi} 
                        - c_\xi v^l_{1,\xi}),\\
  E_l v^l_{2,\xi} = &~& J_2(v^l_{2,\xi-1} + v^l_{2,\xi+1}) 
                        + \underline{\mathcal{U}}_2
                        v^l_{2,\xi} + U_{22} d^{*2}_\xi u^l_{2,\xi} \nonumber\\
                    &-& U_{12} d^{*}_\xi(c_\xi v^l_{1,\xi} 
                        - c^{*}_\xi u^l_{1,\xi}),
 \end{eqnarray}
 \label{hfb_eq_2sp}                 
\end{subequations}
where $\mathcal{U}_1 = 2 U_{11} (n^{c}_{1\xi} + \tilde{n}_{1\xi}) 
+ U_{12} (n^{c}_{2\xi} + \tilde{n}_{2\xi}) + (\epsilon^{(1)}_\xi - \mu_1)$,
$\mathcal{U}_2 = 2 U_{22} (n^{c}_{2\xi} + \tilde{n}_{2\xi}) 
+ U_{12} (n^{c}_{1\xi} + \tilde{n}_{1\xi}) + (\epsilon^{(2)}_\xi - \mu_2)$ with
$\underline{\mathcal{U}}_k = -\mathcal{U}_k$. To solve the above eigenvalue 
equation, we use a basis set of on-site Gaussian wave functions, and define 
the quasiparticle amplitude as linear combination of the basis functions. The 
condensate and noncondensate densities are then computed through the 
self-consistent solution of Eqs.~(\ref{dnls2d}) and (\ref{hfb_eq_2sp}). The 
noncondensate atomic density 
at the $\xi$th lattice site is
\begin{equation}
 \tilde{n}_{k\xi} = \sum_l \left[ (|u^l_{k\xi}|^2 + |v^l_{k\xi}|^2)N_0(E_l) 
                   + |v^l_{k\xi}|^2 \right],
\end{equation}
where $N_0(E_l) = (e^{\beta E_l} - 1)^{-1}$ with $\beta = (k_{B}T)^{-1}$ is 
the Bose-Einstein distribution factor of the $l$th quasiparticle mode with 
energy $E_l$ at temperature $T$. The last term in $\tilde{n}_{k\xi}$ is 
independent of the temperature, and hence, represents the quantum fluctuations 
of the system. To examine the role of temperature we define the miscibility of 
the condensates in terms of the overlap integral
\begin{equation}
\Lambda = \frac{\left[\int n_1(\mathbf r) n_2(\mathbf r) d\mathbf{r}\right]^2}
               {\left[\int n^2_1(\mathbf r) d\mathbf{r} \right] 
                \left[\int n^2_2(\mathbf r) d\mathbf{r} \right]}.
\end{equation}
Here, $n_k (\mathbf r)$ is the total density of $k$th condensate at position 
$\mathbf r \equiv (x,y)$. If the two condensates of the TBEC have complete 
overlap to each other then the system is in miscible phase with $\Lambda=1$, 
whereas for the completely phase-separated case $\Lambda=0$. Using $\Lambda$ as
a measure we identify the miscible and immiscible domain as a function of the 
temperature. As we use the coupled DNLSEs to describe the TBEC, our study is 
valid deep within the superfluid domain, and the mean-field description would 
begin to deviate from the true results near the superfluid-Mott-insulator phase
transition. In this regime a full quantum description~\cite{klaiman_17} would 
be the appropriate method, and the same applies to probing the nature of the 
quantum phase transition~\cite{krauth_92,kashurnikov_96,capello_07,ramanan_09}.
It is well established that for some parameter regimes, TBECs in optical 
lattices can either be superfluid phase of both the species, or superfluid 
phase of one species coexisting with the Mott insulator phase of the 
other~\cite{kashurnikov_02,isacsson_05,barankov_07,mitra_08}.


\subsection{Field-field correlation function}
 To define a measure of the coherence in the condensate we introduce the 
first-order correlation function $g^{(1)}_{k} (\mathbf r, \mathbf r')$, which 
can be expressed as expectations of product of field operators at different 
positions and times~\cite{glauber_63,naraschewski_99,bezett_08,bezett_12}. 
These are normalized to obtain unit modulus in the case of perfect coherence or
a system consisting of only condensate atoms. Here, we restrict ourselves to 
ordered spatial correlation functions at a fixed and equal time. In terms of 
the quantum Bose field operator $\hat{\Psi}_{k}$ the first-order spatial 
correlation function is
\begin{equation}
 g^{(1)}_{k} (\mathbf r, \mathbf r') = \frac{\langle
    \hat{\Psi}^{\dagger}_{k}(\mathbf r)\hat{\Psi}_{k}(\mathbf r')\rangle}
    {\sqrt{{\langle \hat{\Psi}^{\dagger}_{k}(\mathbf r)
     \hat{\Psi}_{k}(\mathbf r) \rangle} {\langle 
     \hat{\Psi}^{\dagger}_{k}(\mathbf r')\hat{\Psi}_{k}(\mathbf r') 
     \rangle}}},
\end{equation}
where $\langle\cdots\rangle$ represents thermal average. It is important to 
note that the local first-order correlation function is equal to the density, 
{\rm i.e.} $g^{(1)}_{k}(\mathbf r, \mathbf r) = n_k(\mathbf r)$. The expression
of $g^{(1)}_{k} (\mathbf r, \mathbf r')$ can also be written in terms of 
condensate, and noncondensate density correlations as
\begin{equation}
 g^{(1)}_{k} (\mathbf r, \mathbf r') = \frac{n^{c}_k(\mathbf r, \mathbf r')
               + \tilde{n}_k(\mathbf r, \mathbf r')}
                 {\sqrt{n_k(\mathbf r) n_k(\mathbf r')}},
 \label{corr_eq}
\end{equation}
where 
\begin{eqnarray*}
 n^{c}_k(\mathbf r,\mathbf r') &=& \psi^{*}_k(\mathbf r) \psi_k(\mathbf r'), \\
 \tilde{n}_k (\mathbf r, \mathbf r') &=& \sum_l \big[\left\{
                       u^{*l}_{k}(\mathbf r) u^{l}_{k}(\mathbf r') 
                       + v^{*l}_{k}(\mathbf r) v^{l}_{k}(\mathbf r') 
                       \right\}N_0(E_l) \\
                     &&+ v^{*l}_{k}(\mathbf r) v^{l}_{k}(\mathbf r') \big], \\
  n_{k}(\mathbf r) &=& n^{c}_k(\mathbf r) + \tilde{n}_k (\mathbf r) 
\end{eqnarray*}
are the condensate density correlation, noncondensate density correlation, and 
total density of the $k$th species, respectively. In the above expressions,  
$n^{c}_k(\mathbf r,\mathbf r')$ and $\tilde{n}_k (\mathbf r, \mathbf r')$ are 
obtained by expanding the complex amplitudes ($c_{\xi},d_{\xi}$) and the 
quasiparticle amplitudes ($u^l_{k,\xi},v^l_{k,\xi}$) in the localized Gaussian 
basis. At $T=0$ K, the entire condensate cloud has complete coherence, and 
therefore $g^{(1)}_{k}=1$ within the condensate region. In TBECs, the 
transition from phase-separated to the miscible domain at $T \neq 0$ has 
characteristic signatures in the spatial structure of 
$g^{(1)}_{k} (\mathbf r, \mathbf r')$. 


\subsection{Numerical methods}
 To solve the coupled DNLSEs, Eqs.~(\ref{dnls2d}), we scale and rewrite the 
equations in the dimensionless form. For this we choose the characteristic 
length scale as the lattice constant $a=\lambda_L/2$ with $\lambda_{L}$ as 
the wavelength of the laser which creates the lattice potential. Similarly, 
the recoil energy $E_R = \hbar^2k_L^2/2m$ with $m$ is the atomic mass of the 
species and $k_L=2\pi/\lambda_L$ is chosen as the energy scale of the system. 
We use fourth-order Runge-Kutta method to solve these equations for zero as 
well as finite temperatures. To initiate the iterative steps to solve the 
equations an appropriate initial guess value of $c_{\xi}$ and $d_{\xi}$ are 
chosen. For the present work we chose the values corresponding to the 
side-by-side profile as it gives quasiparticle energies which are real and not 
complex. This is important as this shows that the solution we obtain is a 
stable one, and not a metastable one. The stationary ground-state wave-function
of the TBEC is obtained through imaginary-time propagation. In the 
tight-binding limit, the width of the orthonormalized Gaussian basis functions 
localized at each lattice site is $0.3a$. Furthermore, to study the 
quasiparticle excitation spectrum, we cast Eqs.~(\ref{hfb_eq_2sp}) as matrix 
eigenvalue equation, and diagonalize the matrix using the routine ZGEEV from 
the LAPACK library~\cite{anderson_99}. For finite temperature computations, to 
take into account the thermal fluctuations, we solve the coupled equations 
Eqs.~(\ref{dnls2d}) and Eqs.~(\ref{hfb_eq_2sp}) self-consistently. The solution
of the DNLSEs is iterated until it satisfies the convergence  criteria in terms
of the number of condensate and noncondensate atoms. In general, the 
convergence is not smooth, and most of the time we encounter severe 
oscillations in the number of atoms. To remedy these oscillations and attain 
convergence, we damp the solution using the successive over- (under-) 
relaxation technique while updating the condensate (noncondensate) atoms. 
Thus, the new solutions after an iteration cycle (IC) are
\begin{subequations}
 \begin{eqnarray}
  c^{\rm new}_{\xi,\rm IC} = r^{\rm ov} c_{\xi,\rm IC} 
                                 + (1 - r^{\rm ov}) c_{\xi,\rm IC-1},  \\
  d^{\rm new}_{\xi,\rm IC} = r^{\rm ov} d_{\xi,\rm IC} 
                                 + (1 - r^{\rm ov}) d_{\xi,\rm IC-1},  \\ 
  \tilde{n}^{\rm new}_{k\xi,\rm IC} = r^{\rm un} \tilde{n}_{k\xi,\rm IC} 
                                 + (1 - r^{\rm un}) \tilde{n}_{k\xi,\rm IC-1},
 \end{eqnarray}
\end{subequations}
where $r^{\rm ov} > 1$ ($r^{\rm un} < 1$) is the over (under) relaxation 
parameter. The choice of  $r^{\rm ov}$ and $r^{\rm un}$ depend on the 
temperature and interaction parameters. In general, our observation is that 
the oscillations are more prominent at higher temperatures, and hence, lower 
values of $r^{\rm ov}$ and $r^{\rm un}$ must be chosen. This in turn implies 
that it takes larger number of iterations to get converged solutions at higher 
temperatures.


\begin{figure}[ht]
 {\includegraphics[width=8.5cm] {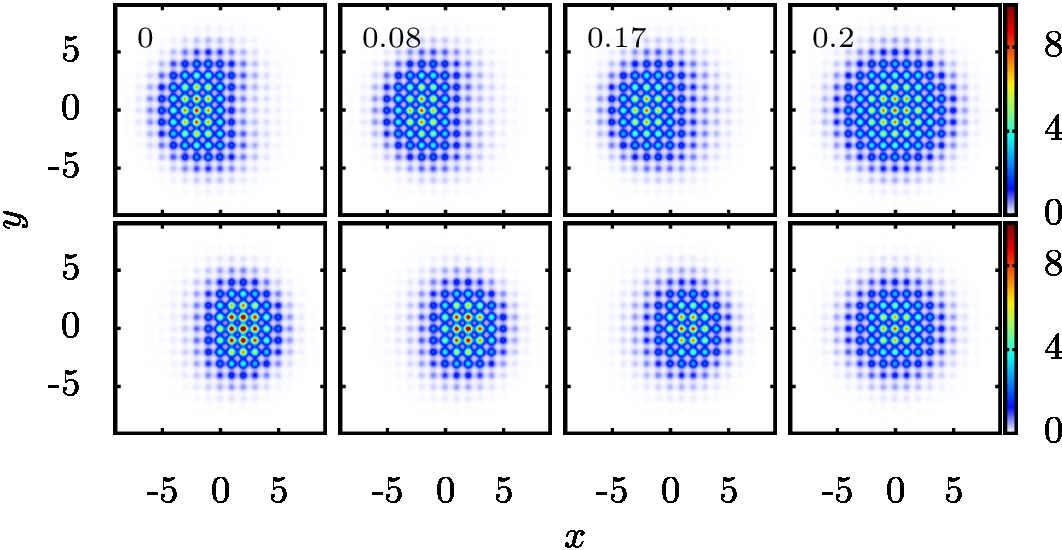}}
  \caption{The density distribution for the condensate atoms of 
           $^{87}$Rb -$^{85}$Rb TBEC as a function of the temperature $T/T_c$.
           The density profiles of the $^{87}$Rb (upper panel) and $^{85}$Rb 
           species (lower panel) are shown for $T/T_c = 0, 0.08, 0.17$, and 
           $0.2$. In the phase-separated domain, the condensate density has 
           side-by-side geometry at zero temperature, and as temperature is 
           increased, there is a transition to miscible domain or the densities
           completely overlap at $T_{\rm ch} = 0.185 T_c$. Here $x$ and $y$ are
           measured in units of the lattice constant $a$.}  
  \label{den_cond_rb}
\end{figure}
\begin{figure}[ht]
 {\includegraphics[width=8.5cm] {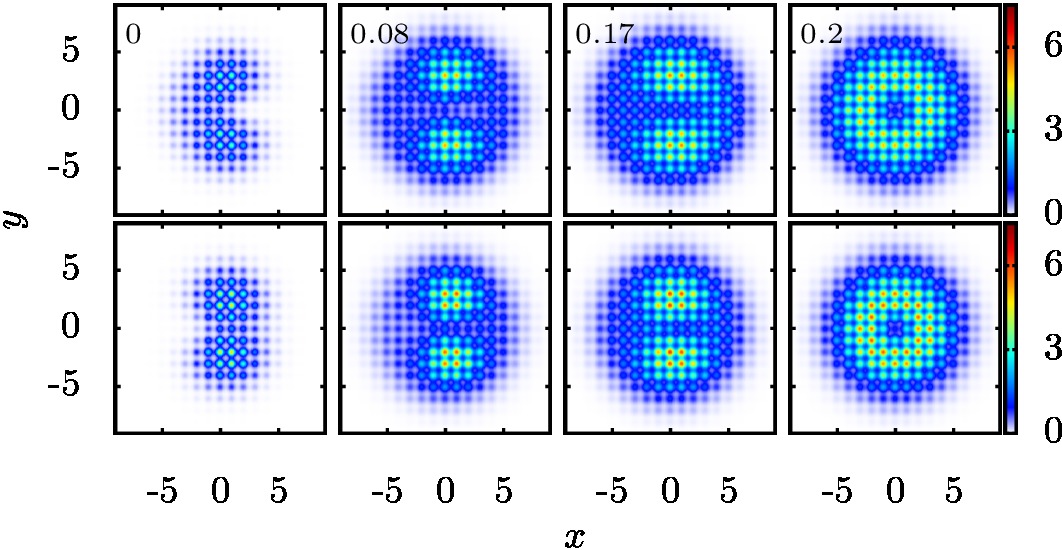}}
  \caption{The density distribution for the noncondensate atoms of 
           $^{87}$Rb -$^{85}$Rb TBEC as a function of the temperature $T/T_c$. 
           The noncondensate density of the $^{87}$Rb (upper panel) and 
           $^{85}$Rb species (lower panel) are shown for $T/T_c = 0, 0.08, 
           0.17$, and $0.2$. The noncondensate atoms which are localized at the
           edges, acquire rotational symmetry in the miscible phase, which 
           happens at $T_{\rm ch} = 0.185 T_c$ as the temperature is increased.
           Here $x$ and $y$ are measured in units of the lattice constant $a$.}
  \label{den_nc_rb}
\end{figure}

\section{Results and discussions}
\label{results}
 To examine the effects of thermal fluctuations on the quasiparticle spectra 
we consider the $^{87}$Rb -$^{85}$Rb TBEC with $^{87}$Rb labeled as species 
$1$ and $^{85}$Rb labeled as species $2$. The radial trapping frequencies of 
the harmonic potential are 
$\omega_x = \omega_y = \omega_{\perp} = 2\pi\times 50$ Hz with the anisotropy 
parameter $\omega_z/\omega_{\perp} = 20.33$, and these parameters are chosen 
based on the experimental work of Gadway and collaborators~\cite{gadway_10} on 
the TBEC of two hyperfine states of $^{87}$Rb in optical lattices. It is 
important to note that we consider equal background trapping potential for 
species $1$ and $2$. We emphasized here that, the results are equally 
applicable to the case of the TBEC consisting of two hyperfine states of 
$^{87}$ Rb, however, we have chosen $^{87}$Rb -$^{85}$Rb to highlight that the 
small mass difference have no influence on the geometry of the ground state. 
The laser wavelength used to create the 2D lattice potential and the lattice 
depth are $\lambda_L = 1064$ nm and $V_0 = 5 E_R$, respectively. We then take 
the total number of atoms as $N_1 = N_2 = 100$ confined in a 
$40\times40$ quasi-2D lattice system. It must be mentioned that the number 
of lattice sites considered much larger than the spatial extent of the 
condensate cloud. Albeit the computations require longer time with the larger 
lattice size, we chose it to ensure that the spatial extent of the thermal 
component is confined well within the lattice considered. The tunneling matrix 
elements are $J_1 = 0.66 E_R$ and $J_2 = 0.71 E_R$, which correspond to an 
optical lattice potential with a depth of $5 E_R$. The intraspecies and 
interspecies on-site interactions are set as $U_{11} = 0.07 E_R$, 
$U_{22} = 0.02 E_R$ and $U_{12} = 0.15 E_R$, respectively. For this set of 
parameters the ground-state density distribution of $^{87}$Rb -$^{85}$Rb TBEC 
is phase-separated with side-by-side geometry. This is a symmetry-broken 
profile where one species is placed to the left and other to the right of the 
trap center along $y$-axis. The evolution of the ground state from miscible to 
the side-by-side density profile due to decrease in the $U_{22}$ is reported 
in our previous work~\cite{suthar_16}. In the present work, we demonstrate the
role of temperature in the phase-separated domain of the binary condensate.
\begin{figure}[ht]
 {\includegraphics[width=8.5cm] {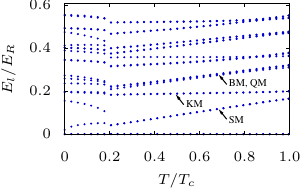}}
  \caption{The evolution of the excitation energies of the low-lying 
           quasiparticle modes as a function of the temperature in 
           $^{87}$Rb -$^{85}$Rb TBEC. The slosh and some of the other higher 
           energy modes become degenerate at $T_{\rm ch} = 0.185 T_c$, where 
           the density distribution is transformed from phase-separated to the 
           miscible profile. In the plot, the slosh mode (SM), Kohn mode (KM), 
           breathing mode (BM), and quadrupole mode (QM) are marked by the 
           black arrows. Here, the excitation energy $E_l$, and the temperature
           $T$ are scaled with respect to the recoil energy $E_R$, and the 
           critical temperature $T_c$ of the $^{87}$Rb species, respectively.}
  \label{mode_rb}
\end{figure}


\subsection{Zero temperature}
 At zero temperature, in the phase-separated domain, the energetically 
preferable ground state of TBEC is the side-by-side geometry, which is reported
in our previous work~\cite{suthar_16}. Unlike in one-dimensional 
system~\cite{suthar_15} in quasi-2D system the presence of the quantum 
fluctuations does not alter the ground state. For the parameters chosen 
$^{87}$Rb -$^{85}$Rb TBEC is phase separated, and the overlap integral has the 
value $\Lambda = 0.10$. The density distributions of the condensate and 
noncondensate atoms of the two species at zero temperature is shown in 
Fig.~\ref{den_cond_rb} and Fig.~\ref{den_nc_rb}. This is a symmetry broken 
side-by-side geometry with noncondensate atoms more localized at the edges of 
the condensate along $y$-axis. 
\begin{figure}[ht]
 {\includegraphics[width=8.5cm] {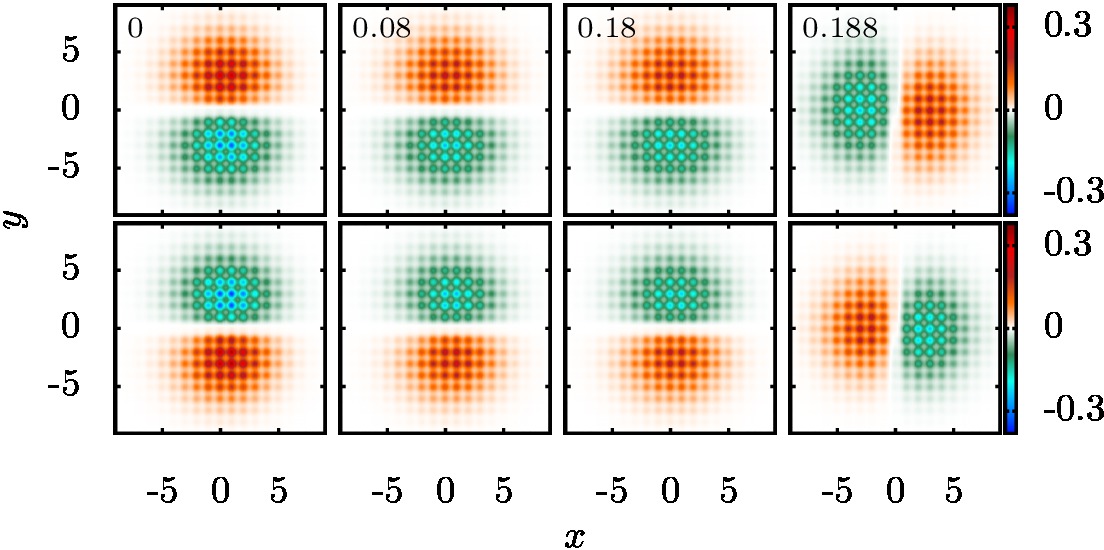}}
  \caption{The quasiparticle mode function of the first excited mode 
           (slosh mode) as a function of the temperature for the 
           $^{87}$Rb -$^{85}$Rb TBEC. The mode function corresponding to the 
           $^{87}$Rb and $^{85}$Rb species are shown in the upper and lower 
           panel, respectively. The slosh mode is an out-of-phase mode where 
           the density flow of the two species are in opposite directions. As 
           the TBEC acquires rotational symmetry at $T_{\rm ch} = 0.185 T_c$, 
           the slosh mode is rotated by an angle $\pi/2$ for 
           $T/T_c\geqslant 0.185$. The value of $T/T_c$ is shown at the upper 
           left corner of each plot in the upper panel. The spatial coordinate 
           $x$ and $y$ are in units of the lattice constant $a$.}
  \label{mode_fn1}
\end{figure}
\begin{figure}[ht]
 {\includegraphics[width=8.5cm] {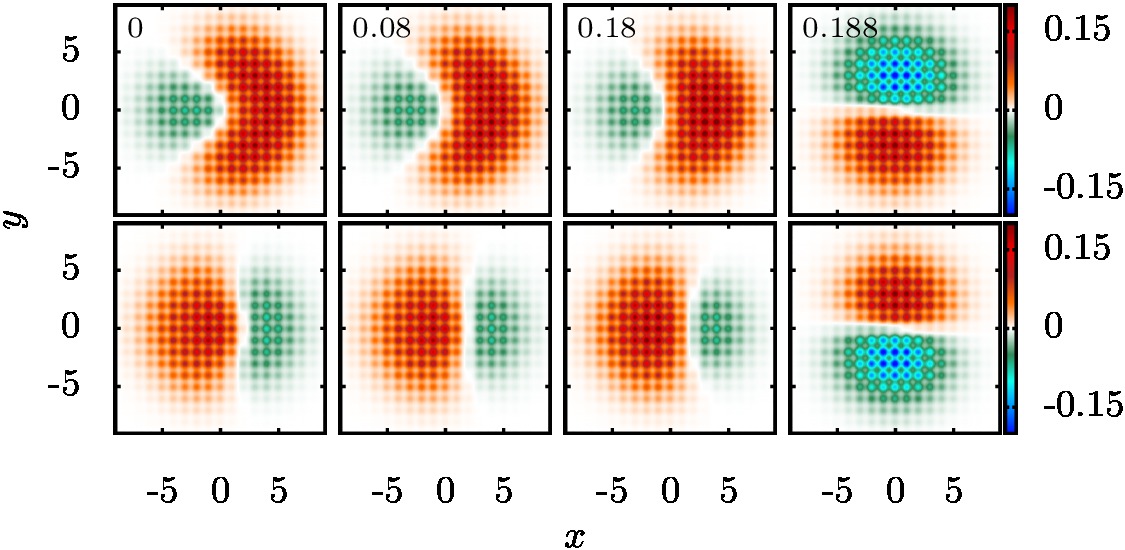}}
  \caption{The quasiparticle mode function corresponding to the second excited 
           mode (slosh mode), which becomes degenerate with the first excited 
           mode for $T/T_c\geqslant 0.185$. The mode function of the 
           $^{87}$Rb and $^{85}$Rb species are shown in the upper and lower 
           panel, respectively. The value of $T/T_c$ is shown at the upper left
           corner of each plot in the upper panel. Here $x$ and $y$ are in 
           units of the lattice constant $a$.}
  \label{mode_fn2}
\end{figure}
\begin{figure}[ht]
 {\includegraphics[width=8.5cm] {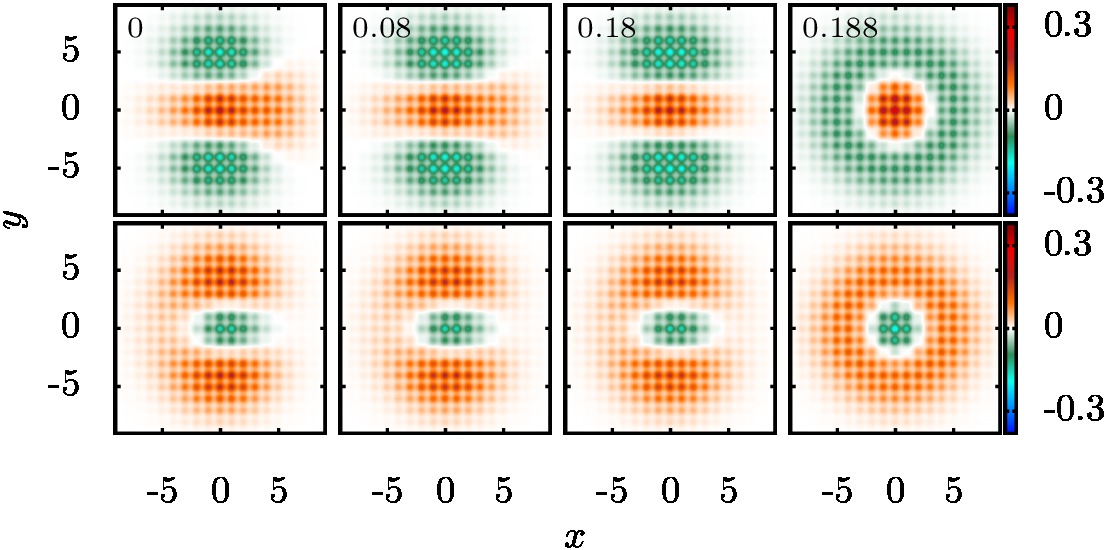}}
  \caption{The quasiparticle mode function corresponding to the interface
           mode in the phase-separated domain of $^{87}$Rb -$^{85}$Rb TBEC as 
           a function of the temperature. This is an out-of-phase mode and the
           mode function is more prominent at the interface region between the 
           condensates. For $T/T_c\geqslant 0.185$, when the TBEC acquires the 
           rotational symmetry, this mode is transformed into the out-of-phase 
           breathing mode, where the mode functions are radially symmetric. The
           value of $T/T_c$ is shown at the upper left corner of each plot in 
           the upper panel. Here $x$ and $y$ are in units of the lattice 
           constant $a$.}        
   \label{mode_fn14}
\end{figure}


\subsection{Finite temperatures}
 At $T\neq0$, in addition to the quantum fluctuations, which are present at
the zero temperature, the thermal cloud also contribute to the noncondensate 
density. As shown in Figs.~\ref{den_cond_rb} and~\ref{den_nc_rb}, at 
$T/T_c = 0.08$, the condensate density profiles of both the species begin to 
overlap, or in other words, the two species are partly miscible. This is also 
evident from the value of $\Lambda=0.16$, which shows a marginal increase
compared to the value of 0.10 at zero temperature. In the figures, the 
temperature is defined in units of the critical temperature $T_c$ of $^{87}$Rb 
atoms, which for the parameters considered is $338$ nK based on our finite
temperature computations. This value of $T_c$ is consistent with the analytic 
expression for ideal Bose gas in optical lattices~\cite{baillie_09}
\begin{equation}
 T_c  = \frac{m\omega^2 a^2}{2\pi k_B} 
          \left[\frac{N_k}{\zeta(3/2)}\right]^{2/3},
\end{equation} 
where $\omega$ is the geometric mean of the three oscillator frequencies, $N_k$
is the number of atoms of $k$th species and $\zeta(3/2) = 2.612$ is the Riemann
zeta function. In the presence of the harmonic confinement, the repulsive 
interatomic interaction reduce the density at the trap center and hence 
decreases $T_c$~\cite{baillie_09}. Upon further increase in temperature, at 
$T/T_c = 0.18$, $\Lambda = 0.36$, this indicates an increase in the miscibility
of the two species. Another important feature at $T/T_c = 0.08$ and $0.18$ is 
the localization of the noncondensate atoms at the interface. This is due to 
repulsion from the condensate atoms, and lower thermal energy which is 
insufficient to overcome this repulsion energy. The transition to miscible 
domain occurs when the temperature exceeds the characteristic temperature
\begin{equation}
  T_{\rm ch} \approx \frac{\sqrt{n_{1\rm max} n_{2\rm max}}~U_{12}}{k_B}, 
\end{equation}
where $n_{k\rm max}$ is the maximum density of the $k$th species. At higher 
temperatures, the extent of overlap between the condensate density profiles 
increases, and TBEC is completely miscible at 
$T_{\rm ch} = 0.185~T_c\approx 63$ nK. This is reflected in the value of 
$\Lambda = 0.95$, and the condensate as well as the noncondensate densities 
acquire rotational symmetry. The $T_{\rm ch}$ at which this transition occurs 
correspond to the thermal energy $k_B T_{\rm ch}=0.72 E_R$, which is comparable
to the the interspecies interaction energy of $0.66 E_R$. Albeit, we discuss in
detail the results for the parameters mentioned earlier, we find similar trends
in the immiscible-miscible transition for different values of $J$'s and $U$'s. 
As to be expected the only change is that the $T_{\rm ch}$ is lowered with 
higher $J$. This is due to the higher kinetic energy associated with higher 
$J$, and hence the atoms require less thermal energy to overcome the 
interspecies repulsion energy for transition to the miscible phase. In terms of
the interaction energies, the lower value of $U_{kk}$ and higher value of 
$U_{12}$ increase the $T_{\rm ch}$ of the TBEC. 

The transition from the phase-separated into miscible domain can further be 
examined from the evolution of the quasiparticle modes as a function of the 
temperature. The evolution of the few low-lying mode energies with temperature 
is shown in Fig.~\ref{mode_rb}, where the temperature is defined in units of 
$T_c$. It is evident from the figure that there are mode energy bifurcations 
with the increase in the temperature. These are associated with the restoration
of rotational symmetry when the TBEC is rendered miscible through an increase 
in temperature. 

 As to be expected the two lowest energy modes are the zero energy or the 
Goldstone modes, which are the result of the spontaneous symmetry breaking 
associated with the condensation. In the phase-separated domain, these modes 
correspond to one each for each of the species. The first two excited modes are
the non-degenerate Kohn or slosh modes of the two species, and these remain 
non-degenerate in the domain $T < T_{\rm ch}$. The structure of these modes are
shown in Figs.~\ref{mode_fn1} and~\ref{mode_fn2}. When $T\geqslant T_{\rm ch}$ 
as the TBEC acquires a rotational symmetry, the slosh modes becomes degenerate 
with $\pi/2$ rotation. A key feature in the quasiparticle mode evolution is 
that the energy of all the out-of-phase mode increases for 
$T\geqslant T_{\rm ch}$, whereas all the in-phase mode remains steady. Here, 
out-of-phase and in-phase means that the amplitudes $u_1$ and $u_2$ of a 
quasiparticle are of different and same phases, respectively. Among the 
low-energy modes, the Kohn mode is in phase whereas the breathing and 
quadrupole modes are out of phase in nature. One unique feature of TBEC in the 
immiscible phase is the presence of interface modes, these have amplitudes 
prominent around the interface region. The existence of these modes is reported
in our previous work~\cite{suthar_16}, and were investigated in other 
works~\cite{ticknor_13,ticknor_14} for TBECs confined in harmonic potential 
alone at zero temperature. As an example, one of the low-energy interface modes
is shown in Fig.~\ref{mode_fn14}. It is evident from the figure that the mode 
is out of phase in nature, and it is transformed into breathing mode in the 
miscible domain when $T\geqslant T_{\rm ch}$. In the miscible domain, the 
breathing mode becomes degenerate with the quadrupole mode, and gains energy. 
The quasiparticles of the miscible domain have well-defined azimuthal quantum 
number, and modes undergo rotations as $T$ is further increased. 
\begin{figure}[h]
 {\includegraphics[width=8.5cm] {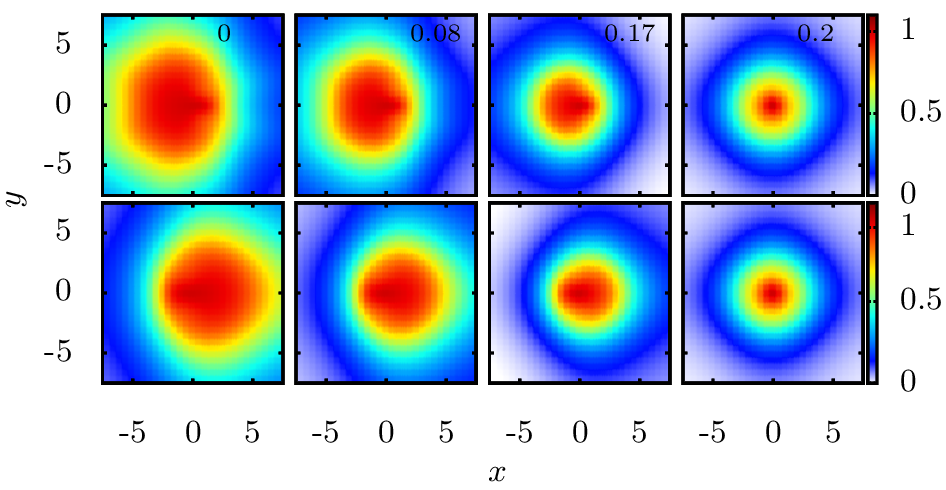}}
  \caption{The normalized first-order spatial correlation function
           $g^{(1)}_{k} (0,\mathbf r)$ for $^{87}$Rb (upper panel) and
           $^{85}$Rb (lower panel) species at $T/T_c = 0, 0.08, 0.17,$
           and $0.2$. Here $x$ and $y$ are measured in units of the lattice 
           constant $a$.}
  \label{corr_fn}
\end{figure}
 To gain additional insights on the immiscible-miscible transition, we consider
other TBECs. In particular, we consider Rb-Cs and Rb-K TBECs confined in 
quasi-2D optical lattices. The details of the parameters chosen and discussion 
are given in the appendix. Starting from the immiscible domain we analyze the 
ground state, and the quasiparticle mode evolution with increase in the 
temperature. Based on the results we observe that the trends in the evolution 
of the low-lying quasiparticle modes with temperature is qualitatively similar 
to the $^{87}$Rb -$^{85}$Rb TBEC. The condensate density profiles also exhibit 
the same trend of transformation from immiscible side-by-side geometry to the 
rotationally symmetric miscible profile. As to be expected, the value of the 
$T_{\rm ch}$ depends on the mass ratio, this is due to the mass dependence of 
interaction energy. In particular, for Rb-Cs and Rb-K TBECs, $T_{\rm ch}$ are 
$0.62~T_c$, and $0.53~T_c$, respectively. The thermal energies corresponding to 
these temperatures are $2.15 E_R$ and $2.80 E_R$, respectively. These are 
comparable to the interaction energies of the TBECs, which are $1.97 E_R$ and 
$2.84 E_R$, respectively. Here $T_c$ is the critical temperature of the 
condensation for the species with lower value. In addition to the atomic mass 
of the condensates, as mentioned earlier, the immiscible-miscible transition 
also depends on the lattice parameters $U$'s and $J$'s. For these two TBECs 
also we have examined the density distributions with variation in $U$'s and 
$J$'s parameters. We find similar trends in the value of the $T_{\rm ch}$ as in
$^{87}$Rb -$^{85}$Rb TBEC. That is, decrease in $T_{\rm ch}$ with increase in 
$J$, and increase with lower and higher values of $U_{kk}$ and $U_{12}$, 
respectively.
 
 To investigate the spatial coherence of TBEC at equilibrium, we examine the
trends in $g^{(1)}_{k} (0, \mathbf r)$ defined earlier in Eq.~(\ref{corr_eq}), 
and are shown in Fig.~\ref{corr_fn} for various temperatures. As mentioned 
earlier, at zero temperature, $n_k(\mathbf r) \approx n^{c}_{k}(\mathbf r)$ 
have complete phase coherence, and therefore, $g^{(1)}_{k} = 1$ within the 
spatial extent of the condensates, this is shown in Fig.~\ref{corr_fn}. At zero 
temperature or in the limit $\tilde{n}_k \equiv 0$ the correlation function, 
Eq.~(\ref{corr_eq}), resemble a Heaviside function, and the negligible 
contribution from the quantum fluctuations smooth out the sharp edges as 
$g^{(1)}_{k}$ drops to zero. More importantly, in the numerical computations 
this cause a loss of numerical accuracy as it involves division of two small 
numbers in Eq.~(\ref{corr_eq})~\cite{gies_04}. However at finite temperature 
the presence of the noncondensate atoms modify the nature of the spatial 
coherence present in the system. The decay rate of the correlation function 
increases with the temperature, and this is evident from Fig.~\ref{corr_fn}, 
which shows $g^{(1)}_{k} (0, \mathbf r)$ at $T/T_c = 0.08, 0.17$, and $0.2$. 
In addition to this, the transition from phase-separated to the miscible TBEC 
is also reflected in the decay trends of $g^{(1)}_{k} (0, \mathbf r)$.


\section{Conclusions}
\label{conc}
 We have examined the finite temperature effects on the phenomenon of phase 
separation in TBECs confined in quasi-2D optical lattices. As temperature is 
increased the phase-separated side-by-side ground state geometry is 
transformed into miscible phase. For the case of TBEC comprising of $^{87}$Rb 
and $^{85}$Rb, the transformation occurs at the characteristic temperature.
This demonstrates the importance of thermal fluctuations which can make TBECs 
miscible. Based on the present work, in general, the TBEC undergoes transition 
to miscible phase at a characteristic temperature $T_{\rm ch}$. This 
corresponds to the temperature at which the thermal energy overcomes the 
interspecies repulsion energy $\sqrt{n_{1\rm max}n_{2\rm max}}~U_{12}$. The 
other key observation is that the transition from phase-separated domain to 
miscible domain is associated with a change in the nature of the quasiparticle 
energies. The low-lying out-of-phase mode, in particular, the slosh mode 
becomes degenerate and increase in energy. On the other hand, the in-phase 
mode, such as Kohn mode, remains steady as temperature ($T<T_c$) is increased. 
The interface modes, which are unique to the phase-separated domain, in 
addition to change in energy are geometrically transformed into rotationally 
symmetric breathing modes in the miscible domain. The temperature driven 
immiscible to the miscible transition is also evident in the profile of the 
correlation functions.


\begin{acknowledgments}
 We thank Arko Roy, S. Gautam, S. Bandyopadhyay and R. Bai for useful 
discussions. The results presented in the paper are based on the computations 
using Vikram-100, the 100TFLOP HPC Cluster at Physical Research Laboratory, 
Ahmedabad, India. 

\end{acknowledgments}

\section*{Appendix}
 Here, we provide brief descriptions of the computations pertaining to the
Rb-Cs and Rb-K TBECs confined in quasi-2D optical lattices. 

\subsection{$^{87}$Rb-$^{133}$Cs TBEC}
  We consider $^{87}$Rb-$^{133}$Cs TBEC containing $100$ atoms of each species 
confined in a $40\times 40$ quasi-2D optical lattices with the wavelength of 
the laser beams as $1064$ nm. The lower number of atoms are chosen to improve 
the convergence of finite temperature computations, and at the same time
it is sufficient to provide a good description of the superfluid phase of the 
TBECs. The radial trapping frequencies of the external harmonic trapping 
potential are $\omega_x = \omega_y = \omega_{\perp} = 2\pi\times 50$ Hz with the 
anisotropy parameter as $20.33$~\cite{gadway_10}. The tunneling matrix elements 
are $J_1 = 0.66 E_R$ and $J_2 = 1.70 E_R$ corresponding to the depth of optical 
lattice $V_0 = 5 E_R$. The lattice depth are considered such that the 
tight-binding limit, $V_0 \gg \mu_k$, is valid. The large difference in the 
values of $J_k$ is due to the large mass difference between the atoms of the 
two species. The intraspecies and interspecies on-site interactions considered 
are $U_{11} = 0.96 E_R$, $U_{22} = 0.42 E_R$ and $U_{12} = 1.2 E_R$. These DNLSE 
parameters are derived from the intra- and interspecies scattering length of 
the species, the trap parameters and the width of the Gaussian basis, which is 
$0.3~a$. At zero temperature, the ground state of the TBEC has side-by-side 
geometry with $\Lambda = 0$~\cite{suthar_16}. Like in the case of 
$^{87}$Rb-$^{85}$Rb, as the temperature of the TBEC is increased ($T<T_c$), the
system is transformed into the miscible phase. In addition, we 
have observed the bifurcation in the energy of the slosh mode, and the mode 
becomes degenerate with a discontinuity in the quasiparticle spectra at 
$T_{\rm ch} = 0.62~T_c\approx 140$ nK. 

\subsection{$^{87}$Rb-$^{41}$K TBEC}
 In the case of the $^{87}$Rb-$^{41}$K TBEC, the wavelength of the laser beams 
and the number of atoms are considered same as in the case of 
$^{87}$Rb-$^{133}$Cs TBEC. The radial trapping frequencies are 
$\omega_x = \omega_y = \omega_{\perp} = 2\pi\times 100$ Hz with the anisotropy 
parameter as $1.40$~\cite{thalhammer_08}. The tunneling matrix elements are 
$J_1 = 0.66 E_R$ and $J_2 = 2.84 E_R$ corresponding to $5 E_R$ lattice depth. 
The intraspecies and interspecies on-site interactions considered are 
$U_{11} = 0.20 E_R$, $U_{22} = 0.06 E_R$ and $U_{12} = 0.60 E_R$. The set of 
parameters are chosen such that the density profile of the TBEC is 
immiscible and has side-by-side geometry at zero temperature. Like in the 
previous cases, the geometry of the TBEC is transformed from side-by-side type 
to the rotationally symmetric overlapping profile and the slosh mode becomes 
degenerate at $T_{\rm ch} = 0.53~T_c\approx 278$ nK.

\bibliography{tbec_2d_temp}{}
\bibliographystyle{apsrev4-1}
\end{document}